\documentclass[fleqn,10pt]{wlscirep}
\usepackage[utf8]{inputenc}
\usepackage[T1]{fontenc}

\title{Temperature dependence of magnetization processes in Sm(Co,Fe,Cu,Zr)$_z$ magnets with different nanoscale microstructures}

\author[1,*]{Leonardo Pierobon}
\author[1,3]{Robin E. Sch\"aublin}
\author[2]{Andr\'as Kov\'acs}
\author[1,3]{Stephan S. A. Gerstl}
\author[1]{Alexander Firlus}
\author[4]{Urs V. Wyss}
\author[2]{Rafal E. Dunin-Borkowski}
\author[1,5]{Michalis Charilaou}
\author[1,*]{J\"org F. L\"offler}

\affil[1]{Laboratory of Metal Physics and Technology, Department of Materials, ETH Zurich, 8093 Zurich, Switzerland}
\affil[2]{Ernst Ruska-Centre for Microscopy and Spectroscopy with Electrons, and Peter Gr\"unberg Institute, Forschungszentrum J\"ulich, 52425 J\"ulich, Germany}
\affil[3]{Scientific Center for Optical and Electron Microscopy, ETH Zurich, 8093 Zurich, Switzerland}
\affil[4]{Arnold Magnetic Technologies, 5242 Birr-Lupfig, Switzerland}
\affil[5]{Department of Physics, University of Louisiana at Lafayette, Lafayette, LA 70504, USA}

\affil[*]{Leonardo Pierobon, leonardo.pierobon@mat.ethz.ch; J\"org F. L\"offler, joerg.loeffler@mat.ethz.ch}


\begin{abstract}
The characteristic microstructure of Sm(Co,Fe,Cu,Zr)$_z$ alloys with SmCo$_5$ cell walls in Sm$_2$Co$_{17}$ cells, all intersected by Zr-rich platelets, makes them some of the best performing high-temperature permanent magnets. Plentiful research has been performed to tailor the microstructure at the nanoscale, but due to its complexity many questions remain unanswered about the effect of the individual phases on the magnetic performance at different temperatures. Here, we explore this mechanism effect for three different Sm(Co,Fe,Cu,Zr)$_z$ alloys by deploying high-resolution magnetic imaging via in-situ transmission electron microscopy and three-dimensional chemical analysis using atom probe tomography. We show that their microstructures differ in terms of SmCo$_5$ cell-wall and Z-phase size and density, as well as the Cu concentration in the cell walls, and demonstrate how these features influence the magnetic domain size and density and thus form different magnetic textures. Moreover, we illustrate that the dominant coercivity mechanism at room temperature is domain-wall pinning and show that magnets with a denser cell-wall network, a steeper Cu gradient across the cell-wall boundary, and thinner Z-phase platelets have a higher coercivity. We also show that the coercivity mechanism at high temperatures is domain-wall nucleation at the cell walls. Increasing the Cu concentration inside the cell walls decreases the transition temperature between pinning and nucleation, significantly decreasing the coercivity with increasing temperature. We therefore provide a detailed explanation of how the microstructure on the atomic to nanoscale directly affects the magnetic performance and provide detailed guidelines for an improved design of Sm(Co,Fe,Cu,Zr)$_z$ magnets.
\end{abstract}
\begin{document}

\flushbottom
\maketitle

\thispagestyle{empty}

\section*{Introduction}

Sm--(Co, Fe, Cu, Zr)$_z$ ($z = 6.7 - 9.1$) alloys are one of the best commercially available permanent magnets for high-temperature applications, owing to their high Curie temperature and large magnetic coercivity \cite{ray1992, gutfleisch2011,mccallum2014}. Their microstructure, which has been nanoengineered over decades of extensive research \cite{hadjipanayis1984, liu1999, Horiuchi2013, duerrschnabel2017}, forms as a result of a carefully designed aging and heat-treatment process. It comprises Sm$_2$Co$_{17}$ cells, SmCo$_5$ cell walls, and Zr-rich platelets, called the Z phase. The platelets are oriented perpendicular to the $c$-axis of the cells, which is also the easy axis of the magnetocrystalline anisotropy. The intertwined structure of the cells with high saturation magnetization and the cell walls with high magnetic anisotropy significantly improves the magnetic properties and makes them highly tunable \cite{kronmueller1996, Skomski2013}. Especially important is the enhancement of coercivity due to domain-wall (DW) pinning, which arises from the difference in the DW energy between the cells and the cell walls because of their different magnetocrystalline anisotropy \cite{gaunt1972, livingston1977, Nagel1979, hadjipanayis1982b, wong1997, skomski1997, Fidler2000, Fidler2004}. This has been well-documented experimentally by Lorentz transmission electron microscopy (LTEM), magnetic force microscopy and Kerr microscopy \cite{fidler1982, hadjipanayis1982, zhang2000, gutfleisch2006, okabe2006, sepehri2017, zhang2018}.

Due to the complexity of the microstructure, many questions about the role of certain microstructural features in the magnetic properties of the magnet remain elusive. It has been proposed that the Z-phase platelets act as diffusion pathways for the formation of SmCo$_5$ cell walls, aiding the segregation of Cu inside the cell walls\cite{rabenberg1982}. However, the magnetic properties of the lamellar Z phase are not well studied, because they differ from their bulk values. A recent study showed that the Z phase has an insignificant magnetic anisotropy compared to the other two phases, and proposed that increasing the thickness of the Z phase would decrease the coercivity of the magnet \cite{pierobon2020}. Furthermore, the size of the Sm$_2$Co$_{17}$ cells affects the coercivity and the remanence, because it determines the chemical composition of the cell walls \cite{streibl2000}. The presence of Cu in the cell walls increases the DW-pinning strength and thus the coercivity \cite{hadjipanayis2000, lectard1994, tang2000}. Simultaneously, this decreases the Curie temperature of the cell walls, as well as the overall remanence, so the effect of Cu and cell size on the high-temperature magnetic properties remains unclear. Finally, magnetization reversal can start with DW nucleation at the cell walls or Z phase \cite{kronmuller2002, goll2004, pierobon2020}, meaning that the magnetization processes in Sm(Co,Fe,Cu,Zr)$_z$ can be either DW pinning or nucleation, or both. In order to fully understand the behavior of Sm(Co,Fe,Cu,Zr)$_z$ magnets, it is necessary to investigate how the aforementioned microstructural features give rise to DW pinning and nucleation, and what determines which one of them is the dominant magnetization process. So far, the limited resolution of magnetic imaging techniques has hindered direct observations and the ability to answer these questions.

Here, we present a systematic study of three Sm(Co,Fe,Cu,Zr)$_z$ magnets with different chemical compositions manufactured using the same heat-treatment process (described in the Methods section). We correlate the magnetic performance of these magnets at different temperatures using high-resolution imaging of their microstructure and magnetic structure. The microstructure has been imaged by energy-dispersive X-ray (EDX) spectroscopy in the scanning TEM (STEM) mode and atom probe tomography (APT), while the magnetic structure has been imaged by in-situ heating LTEM and off-axis electron holography (EH). By correlating these methods and comparing the magnets in detail, we identify the mechanisms of how the microstructure gives rise to complex magnetic structures, such as closed magnetic loops and dense networks of magnetic domains. We also show that the dependence of coercivity on temperature is dominated by DW pinning at low temperatures and nucleation at high temperatures, and explain how these magnetization processes arise from different microstructural features. Finally, we propose how the microstructure should be modified to improve the performance of Sm(Co,Fe,Cu,Zr)$_z$ magnets.

\section*{Results}

The three magnetic samples that were studied, namely Sm\textsuperscript{L}Cu\textsuperscript{L} (low Sm, low Cu), Sm\textsuperscript{L}Cu\textsuperscript{H} (low Sm, high Cu) and Sm\textsuperscript{H}Cu\textsuperscript{H} (high Sm, high Cu), have different chemical compositions, as shown in Table 1. Sm\textsuperscript{L}Cu\textsuperscript{H} has a higher Cu content than Sm\textsuperscript{L}Cu\textsuperscript{L} at the expense of Co. Similarly, Sm\textsuperscript{H}Cu\textsuperscript{H} has a higher Sm content than Sm\textsuperscript{L}Cu\textsuperscript{H} at the expense of Co. The z number, i.e. the ratio of transition metals to Sm, is 7.7 for Sm\textsuperscript{L}Cu\textsuperscript{L} and Sm\textsuperscript{L}Cu\textsuperscript{H}, and 7.3 for Sm\textsuperscript{H}Cu\textsuperscript{H}. 

\begin{table}[h]
\centering
\begin{tabular}{|c||c|c|c|c|c|c|}
\hline
 Sample & Sm [at.\%] & Co [at.\%] & Fe [at.\%] & Cu [at.\%] & Zr [at.\%] & z\\ 
\hline
\hline
 Sm\textsuperscript{L}Cu\textsuperscript{L} & 11.5 & 60.9 & 18.8 & 6.2 & 2.5 & 7.7\\ 
\hline
 Sm\textsuperscript{L}Cu\textsuperscript{H} & 11.6 & 59.4 & 18.8 & 7.7 & 2.5 & 7.7\\ 
\hline
 Sm\textsuperscript{H}Cu\textsuperscript{H} & 12.0 & 58.7 & 18.9 & 7.8 & 2.5 & 7.3\\ 
\hline
\end{tabular}
\caption{Chemical compositions of the samples investigated in this work; $z$ describes the ratio of transition metals to Sm.}
\end{table}

The microstructure of the samples was imaged by STEM EDX spectroscopy on TEM lamellae with the $c$-axis in-plane to ensure identical imaging conditions. The average thickness of the lamellae, as measured by electron energy-loss spectroscopy (EELS), varies between 95 nm and 145 nm, and the thickness variations are bigger within the same lamella than among different lamellae. We imaged locations with different thicknesses and verified that the thickness variation does not affect the measurements qualitatively or quantitatively. Figure 1 reveals that the samples have a typical microstructure of high-performance Sm(Co,Fe,Cu,Zr)$_z$ magnets with Sm$_2$Co$_{17}$ cells, zig-zag SmCo$_5$ cell walls, and the Z phase. It is difficult to perform a statistical analysis of microstructural features, such as the cell-wall separation (the distance between adjacent cell walls with the same direction, $i.e.$ the cell width), the cell-wall thickness, the Z-phase separation (the distance between adjacent Z-phase platelets), and the Z-phase thickness, because the boundaries between the phases are not well-defined and their values vary significantly within the sample. In the following, we discuss the average values measured in the images, and the error analysis is omitted because the measurement error is at least an order of magnitude smaller than the variations within the same sample. Here, it is important to note that TEM measurements are two-dimensional projections of a three-dimensional system. For example, the cell walls and Z-phase platelets may intersect the lamella plane at an angle, which would make their thickness appear bigger than the actual values. However, due to the similar thickness and the same crystal orientation of the lamella, as analyzed in Supplementary Figure 1, these measurements represent a reliable quantitative comparison between the samples.

Zr and Cu concentration maps extracted from the EDX data, indicating the morphology of the Z phase and cell walls, respectively, are shown in Figure 1a,d for Sm\textsuperscript{L}Cu\textsuperscript{L}, Figure 1b,e for Sm\textsuperscript{L}Cu\textsuperscript{H}, and Figure 1c,f for Sm\textsuperscript{H}Cu\textsuperscript{H}. The average Z-phase separation is 44 nm in Sm\textsuperscript{L}Cu\textsuperscript{L}, 67 nm in Sm\textsuperscript{L}Cu\textsuperscript{H}, and 91 nm in Sm\textsuperscript{H}Cu\textsuperscript{H}. Because the amount of Zr is the same in all samples, the difference in the Z-phase separation is compensated by the Z-phase thickness, which is on average 23 nm in Sm\textsuperscript{L}Cu\textsuperscript{L}, 28 nm in Sm\textsuperscript{L}Cu\textsuperscript{H}, and 59 nm in Sm\textsuperscript{H}Cu\textsuperscript{H}. The reason for such a difference in the Z-phase morphology may be due to the fact that the Z phase acts as a diffusion pathway for the formation of the Sm-rich cell walls, thus reflecting the difference in the Sm content between the samples, where the sample with the highest Sm content has the thickest Z phase.

In all samples, the two directions of the zig-zag cell walls are symmetrical with respect to the $c$-axis and make an angle between 55$^\circ$ and 65$^\circ$ with each other. The average cell-wall separation (cell width) is 104 nm in Sm\textsuperscript{L}Cu\textsuperscript{L}, 107 nm in Sm\textsuperscript{L}Cu\textsuperscript{H}, and 81 nm in Sm\textsuperscript{H}Cu\textsuperscript{H}. Since the Sm content and cell-wall separation are almost identical in Sm\textsuperscript{L}Cu\textsuperscript{L} and Sm\textsuperscript{L}Cu\textsuperscript{H} and lower in Sm\textsuperscript{H}Cu\textsuperscript{H}, it can be concluded that the cell-wall separation decreases with increasing amount of Sm, which favors the formation of Sm-rich cell walls. The average cell-wall thickness is 15 nm in Sm\textsuperscript{L}Cu\textsuperscript{L}, 27 nm in Sm\textsuperscript{L}Cu\textsuperscript{H}, and 36 nm in Sm\textsuperscript{H}Cu\textsuperscript{H}. This may be explained by considering three factors. First, increasing the amount of Sm favors the formation of Sm-rich cell walls. Second, Cu segregates inside the cell walls, so a higher Cu content may lead to thicker cell walls. Third, the Z phase acts as a diffusion pathway for the formation of cell walls during the annealing process, so the cell-wall thickness may increase with the Z-phase thickness. This is why Sm\textsuperscript{H}Cu\textsuperscript{H}, with the highest Sm and Cu content and the thickest Z phase, has the thickest cell walls. 

\begin{figure}[h]
\centering
\includegraphics[width=1\columnwidth]{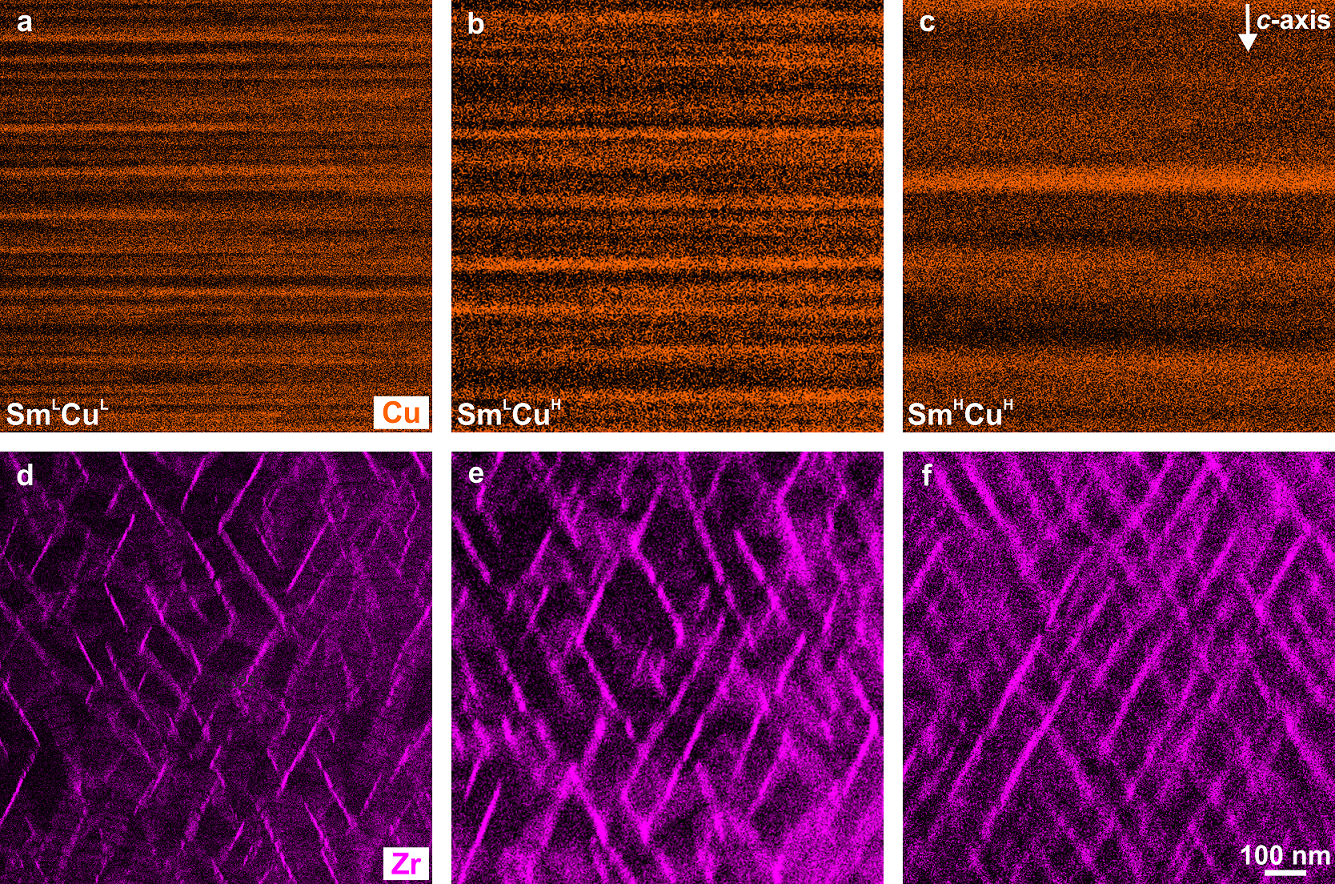} 
\caption{\textbf{EDX chemical analysis.} HRSTEM EDX element maps of Zr and Cu for (\textbf{a, d}) Sm\textsuperscript{L}Cu\textsuperscript{L}, (\textbf{b, e}) Sm\textsuperscript{L}Cu\textsuperscript{H} and (\textbf{c, f}) Sm\textsuperscript{H}Cu\textsuperscript{H}, where bright contrast represents Z-phase platelets and SmCo$_5$ cell walls.}
\end{figure}

The segregation of Cu inside the cell walls plays an important role in DW pinning, but this cannot be reliably quantified in EDX due to the fact that a TEM image is a two-dimensional projection of a three-dimensional structure. For example, if the cell walls intersect the lamella plane at an angle, this may distort an observed Cu concentration profile, making the cell walls appear wider and the Cu concentration lower than the real values. Particularly important is the Cu gradient at the interface between the cells and the cell walls, which can only be resolved with atomic-resolution measurements in three dimensions. For this reason, an APT analysis, as shown in Figure 2, was performed. The three-dimensional reconstruction of the samples (Fig. 2d-f) confirms a typical microstructure with Sm$_2$Co$_{17}$ cells, SmCo$_{5}$ cell walls and a Zr-rich Z phase. Proxigrams (proximity histograms) were extracted from the reconstruction, which show the concentration of individual elements as a function of distance from a predefined interface. Here, the predefined interface corresponds to all surfaces with a Cu concentration of 15\%, which represent the boundaries of the cell walls. This provides statistically relevant information because the concentration profiles are averaged over all the cell walls. The proxigrams shown in Figure 2 reveal that the Cu concentration profile inside the cell walls is bell-shaped. The Cu concentration at 2.5 nm from the interface is 38.6 $\pm$ 0.2\% in Sm\textsuperscript{L}Cu\textsuperscript{L} (Fig. 2a), 42.2 $\pm$ 0.3\% in Sm\textsuperscript{L}Cu\textsuperscript{H} (Fig. 2b) and 34.0 $\pm$ 0.2\% in Sm\textsuperscript{H}Cu\textsuperscript{H} (Fig. 2c). Although Sm\textsuperscript{H}Cu\textsuperscript{H} has the highest Cu content (see Table 1), it has the lowest Cu concentration inside the cell walls, because it has the most numerous and thickest cell walls, over which Cu has to be distributed. Figure 2g compares the Cu profiles of the samples, which were obtained by normalizing the local Cu concentration with respect to its value at 2.5 nm from the interface. Sm\textsuperscript{L}Cu\textsuperscript{H} has the steepest profile, with a gradient of 49.2\%nm\textsuperscript{-1} at the inflection point, followed by Sm\textsuperscript{L}Cu\textsuperscript{L} and Sm\textsuperscript{H}Cu\textsuperscript{H}, with gradients of 43.0\%nm\textsuperscript{-1} and 37.0\%nm\textsuperscript{-1} at the inflection points, respectively. The steepness of the Cu profile has a profound effect on the DW pinning strength, which will be discussed below. 

\begin{figure}[h]
\centering
\includegraphics[width=1\columnwidth]{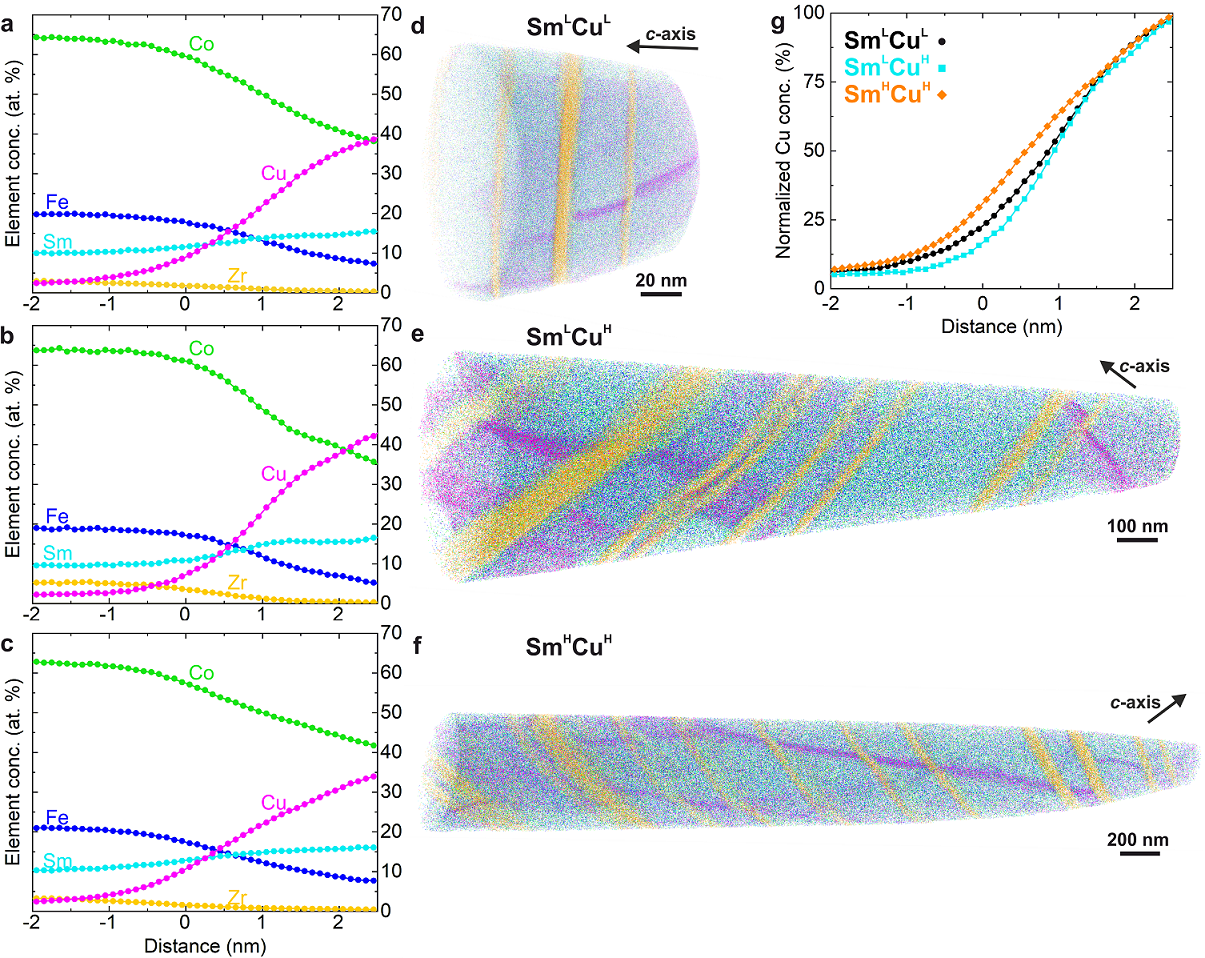} 
\caption{\textbf{APT chemical analysis.} Concentration proxigrams of individual elements with respect to the cell-wall boundaries and APT reconstructions for (\textbf{a, d}) Sm\textsuperscript{L}Cu\textsuperscript{L}, (\textbf{b, e}) Sm\textsuperscript{L}Cu\textsuperscript{H} and (\textbf{c, f}) Sm\textsuperscript{H}Cu\textsuperscript{H}. (\textbf{g}) Cu concentration normalized with respect to its value at 2.5 nm from the cell-wall boundary, revealing the steepness of the Cu gradient for each sample.}
\end{figure}

The APT data was also used to measure the Zr distribution in the Z phase, which is almost identical in all samples, as shown in Supplementary Figure 2. The thickness of the Z phase is 3.1 $\pm$ 0.7 nm for Sm\textsuperscript{L}Cu\textsuperscript{H} and 4.4 $\pm$ 1.5 nm for Sm\textsuperscript{H}Cu\textsuperscript{H}. These data confirm the discussion above that EDX measurements overestimate the thickness of the measured features due to a two-dimensional projection of three-dimensional objects across thick TEM lamellae. Nonetheless, the APT measurements validate the relative comparison obtained by EDX, confirming that the Z phase is thicker in Sm\textsuperscript{H}Cu\textsuperscript{H} than in Sm\textsuperscript{L}Cu\textsuperscript{H}. The thickness was not measured for Sm\textsuperscript{L}Cu\textsuperscript{L}, because the APT reconstruction was not big enough to provide statistically relevant data, as visible in Figure 2d.

The aforementioned nanoscale microstructural features interact with the magnetic texture and the associated magnetization processes because they are comparable in size to the fundamental magnetic length scales in Sm(Co,Fe,Cu,Zr)$_z$ magnets. For example, the theoretical DW width of Sm(Co,Fe,Cu,Zr)$_z$ magnets is between 2.6 and 5.7 nm, and the exchange length, which defines the length scale of magnetization collinearity, is between 5.2 and 8.0 nm \cite{zhao2019}. The magnetic structure of the samples was imaged using LTEM in the Fresnel mode, where DWs create alternating bright and dark (convergent and divergent) contrast at defocus (see Methods). Figure 3 shows that the DWs have a typical zig-zag shape resulting from the pinning at the SmCo$_5$ cell walls. As was the case with the microstructural features, in the following only average values will be discussed due to large variations in the size of the magnetic textures within the same sample. In Sm\textsuperscript{L}Cu\textsuperscript{L} (Fig. 3a), the domains are significantly elongated; their length can span for over a micrometer, and the separation between adjacent DWs, $i.e.$ the width of the domains, is on average 310 nm. In Sm\textsuperscript{L}Cu\textsuperscript{H} (Fig. 3b), the domains are smaller; their length spans between 400 and 600 nm, and the DW separation is 187 nm on average. In Sm\textsuperscript{H}Cu\textsuperscript{H} (Fig. 3c), the domains are significantly smaller, with their length typically spanning between 150 and 250 nm, and the DW separation is 65 nm on average. From this, we can conclude that the DW separation increases with increasing cell-wall separation, so that the DWs are most numerous in Sm\textsuperscript{H}Cu\textsuperscript{H}, just like the SmCo$_5$ cell walls. However, this is not the only factor impacting the DW separation, because Sm\textsuperscript{L}Cu\textsuperscript{L} and Sm\textsuperscript{L}Cu\textsuperscript{H} have approximately the same cell-wall separation, but their DW separation is not the same. This may be explained by considering the concentration of Cu in the cell walls, which will be discussed below. 

\begin{figure}[h]
\centering
\includegraphics[width=1\columnwidth]{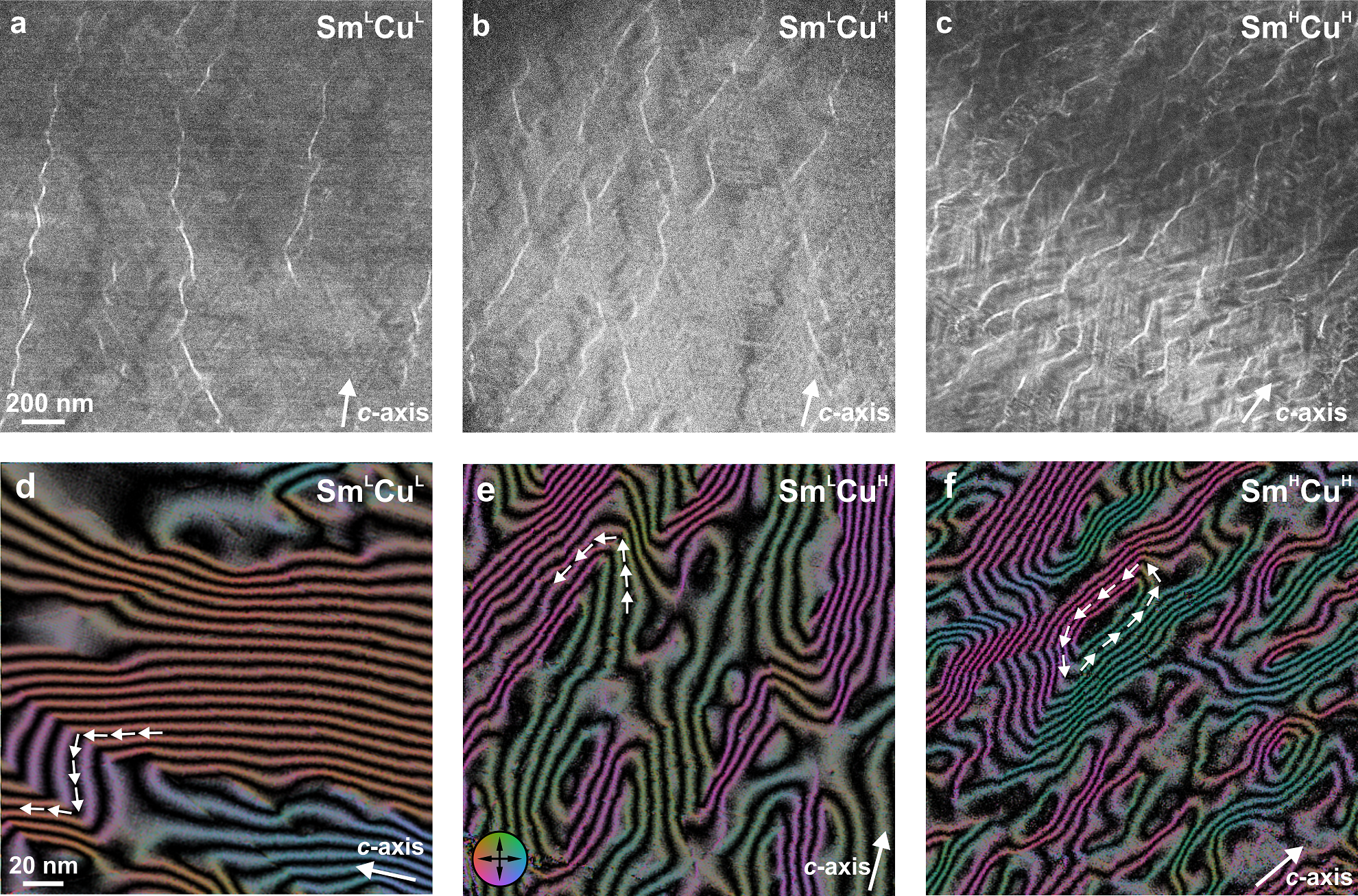}
\caption{\textbf{Magnetic structure.} Fresnel images at 300 $\mu$m overfocus of the thermally demagnetized state of (\textbf{a}) Sm\textsuperscript{L}Cu\textsuperscript{L}, (\textbf{b}) Sm\textsuperscript{L}Cu\textsuperscript{H} and (\textbf{c}) Sm\textsuperscript{H}Cu\textsuperscript{H}, showing the density of DWs in each sample. Magnetic induction maps of (\textbf{d}) Sm\textsuperscript{L}Cu\textsuperscript{L}, (\textbf{e}) Sm\textsuperscript{L}Cu\textsuperscript{H} and (\textbf{f}) Sm\textsuperscript{H}Cu\textsuperscript{H}, with white arrows demonstrating magnetization curling, which in panel f results in a closed magnetic loop. The contour spacing in the maps is $\pi$.}
\end{figure}

The magnetic structure was further analyzed by performing off-axis electron holography (EH) measurements, which provide quantitative information on the in-plane component of the magnetic induction in the lamellae. The magnetic induction maps extracted from EH are shown in Figures 3d-f. The direction of the contours and their color represent the direction of the magnetic field (as indicated by the color bar), while the density of the contour lines represents the magnetic field strength. The contour spacing, $i.e.$ the phase change of the electron wave between adjacent contours, is $\pi$ in all induction maps presented in this paper. Conventionally, only $\pi$ DWs are expected in high-anisotropy magnets \cite{kittel1949}, but the arrows in Figure 3d illustrate that $\pi/2$ DWs are also present in Sm\textsuperscript{L}Cu\textsuperscript{L}. Figure 3e shows that the magnetic structure is even more complex in Sm\textsuperscript{L}Cu\textsuperscript{H}, with a high degree of magnetization curling, resulting in DWs not having a well-defined angle. This is demonstrated with arrows in Figure 3e, where the change in the magnetic field direction across the same DW varies abruptly. The magnetic structure is the most complex in Sm\textsuperscript{H}Cu\textsuperscript{H}, where magnetic curling is so strong that in some cases it leads to closed magnetic loops. An example of such a magnetic loop is marked with arrows in Fig. 3f. The density of the magnetic field lines varies within all samples. Since electron holography can only detect the in-plane component of the magnetic field, a lower density of field lines indicates the presence of a significant out-of-plane component of the magnetic field. One such region can be seen in the top corner of Figure 3d, where the contour spacing is almost flat. Finally, for each sample an external magnetic field of 1.5 T was applied by the objective lens perpendicular to the lamella and no change in the magnetic structure was found, which indicates very strong DW pinning.

The microstructural analysis was directly correlated with the magnetic imaging in order to identify the mechanisms by which the microstructural features and magnetic textures are related. In Figures 4a-c, an EDX Cu concentration map indicating the positions of cell walls (Fig. 4a), an LTEM image indicating the positions of DWs (Fig. 4b), and a magnetic induction map indicating the direction of the magnetic field (Fig. 4c) are compared for the same area in Sm\textsuperscript{L}Cu\textsuperscript{L}. As indicated by white arrows, the DWs are pinned at the SmCo$_5$ cell walls. However, the shape of the DWs does not match exactly the cell-wall morphology, which gives evidence for DWs being situated in the Sm$_2$Co$_{17}$ cells (repulsive pinning). Please note that the wave-like contrast visible in Figure 4b does not come from ion-beam curtaining but is a fringe contrast from the interference because the image was recorded in EH mode. Figures 4d,e compare for the same area in Sm\textsuperscript{H}Cu\textsuperscript{H} an electron-wave amplitude map (extracted from EH), indicating the positions of cell walls with bright contrast (Fig. 4d), and a magnetic induction map indicating the direction of the magnetic field (Fig. 4e). The DWs are pinned at the cell walls and follow them closely. At the intersections between two cell walls, as shown by yellow arrows, the DWs follow one cell wall and then curl in order to keep on following the other cell wall. Such curling can result either in DWs not having a very well defined angle of rotation or in the formation of closed magnetic loops. This finding explains that Sm\textsuperscript{H}Cu\textsuperscript{H} has the most complex magnetic structure (as shown in Figure 3), because it has the densest network of cell walls with numerous intersections. Therefore, by comparing the microstructure and the magnetic structure directly, it has been observed how complex magnetic textures are formed in Sm(Co,Fe,Cu,Zr)$_z$ magnets, enriching our current understanding of magnetic phenomena in permanent magnets.

\begin{figure}[h]
\centering
\includegraphics[width=1\columnwidth]{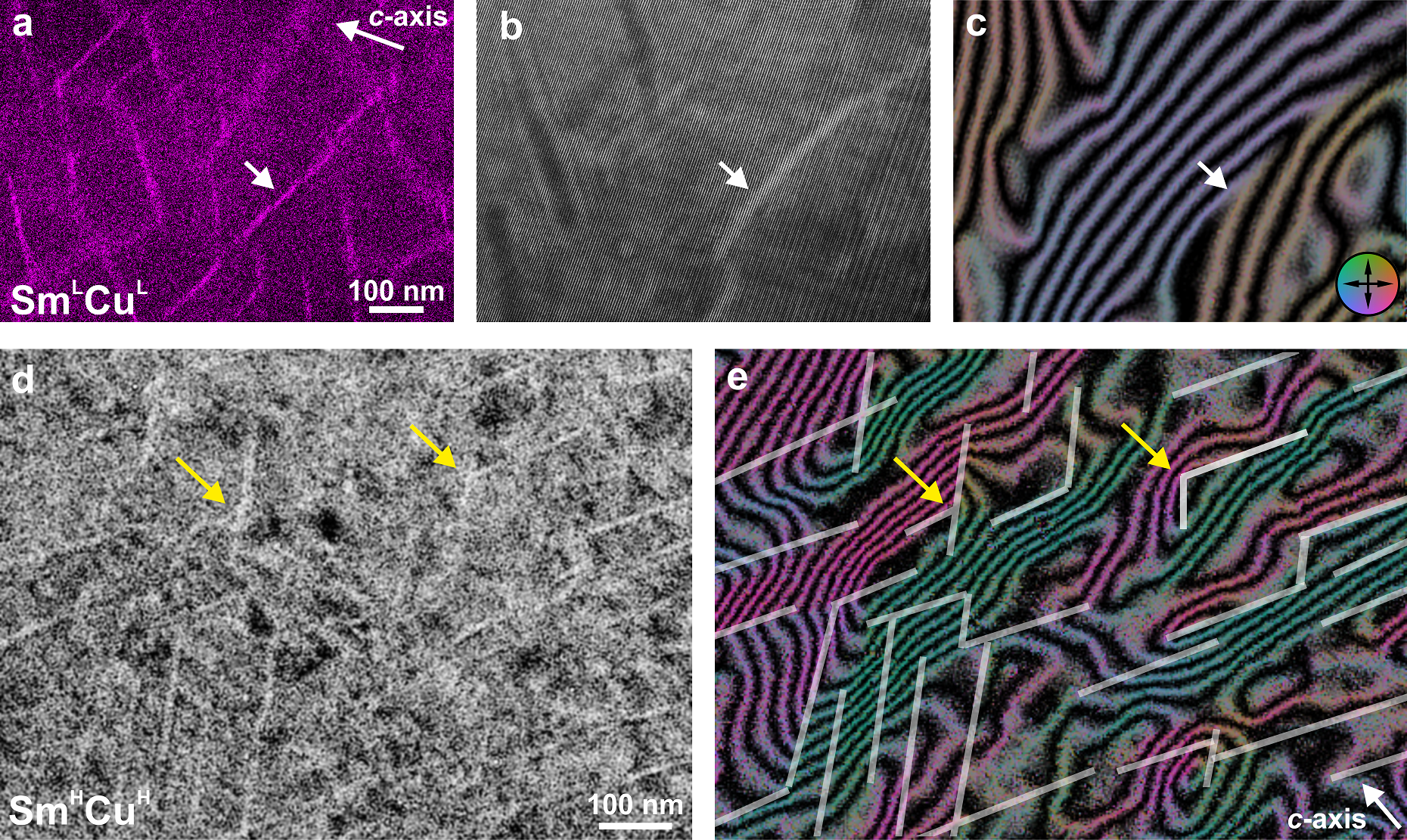}
\caption{\textbf{Correlation of the micro- and magnetic structure.} (\textbf{a}) An EDX concentration map of Cu; (\textbf{b}) a Fresnel image at 200 $\mu$m overfocus and (\textbf{c}) a magnetic induction map of the same area in Sm\textsuperscript{L}Cu\textsuperscript{L}, with white arrows revealing repulsive DW pinning at a cell wall. (\textbf{d}) An electron-wave amplitude map and (\textbf{e}) a magnetic induction map of the same area in Sm\textsuperscript{H}Cu\textsuperscript{H}, exemplifying by yellow arrows that magnetic loops form as a consequence of pinning at cell-wall intersections. The contour spacing in the induction maps is $\pi$. The color wheel indicates the direction of the magnetic induction.}
\end{figure}

In-situ heating LTEM experiments were performed with and without an external magnetic field to investigate the magnetization processes in Sm(Co,Fe,Cu,Zr)$_z$ at high temperatures. There is no observed change in microstructure when the samples are heated from room temperature (RT) to 400 $^{\circ}$C, which is the temperature at which the magnets are stabilized in the annealing process. This means that any observed change in the magnetic structure in this temperature range is due to the change in the magnetic properties and not due to the change in the microstructure. When the samples are heated without an external magnetic field, there is no change in the magnetic structure from RT to 400 $^{\circ}$C. This excludes the possibility for these samples that there is a change from repulsive to attractive pinning with increasing temperature, because the shape of the DWs does not change. However, when the samples are heated in an external magnetic field of 1.5 T perpendicular to the lamella, the magnetic structure changes significantly. Figures 5a-c show LTEM images of the same area in Sm\textsuperscript{L}Cu\textsuperscript{H} at 20 $^{\circ}$C (Fig. 5a), 200 $^{\circ}$C (Fig. 5b) and 250 $^{\circ}$C (Fig. 5c). Some DWs are nucleated and their neighboring DWs are annihilated between 150 $^{\circ}$C and 200 $^{\circ}$C, as indicated by two yellow arrows comparing the same positions in Figures 5a,b. This is therefore evidence of DW nucleation in Sm(Co,Fe,Cu,Zr)$_z$ magnets. As the temperature increases above 200 $^{\circ}$C, the DWs are annihilated (Fig. 5c), because the coercivity decreases and the magnet becomes saturated perpendicular to the lamella plane (parallel to the external magnetic field). At 250 $^{\circ}$C most DWs are annihilated, but some DWs survive even at 300 $^{\circ}$C. 
The full in-situ heating experiment is shown in Supplementary Video 1.

\begin{figure}[h]
\centering
\includegraphics[width=1\columnwidth]{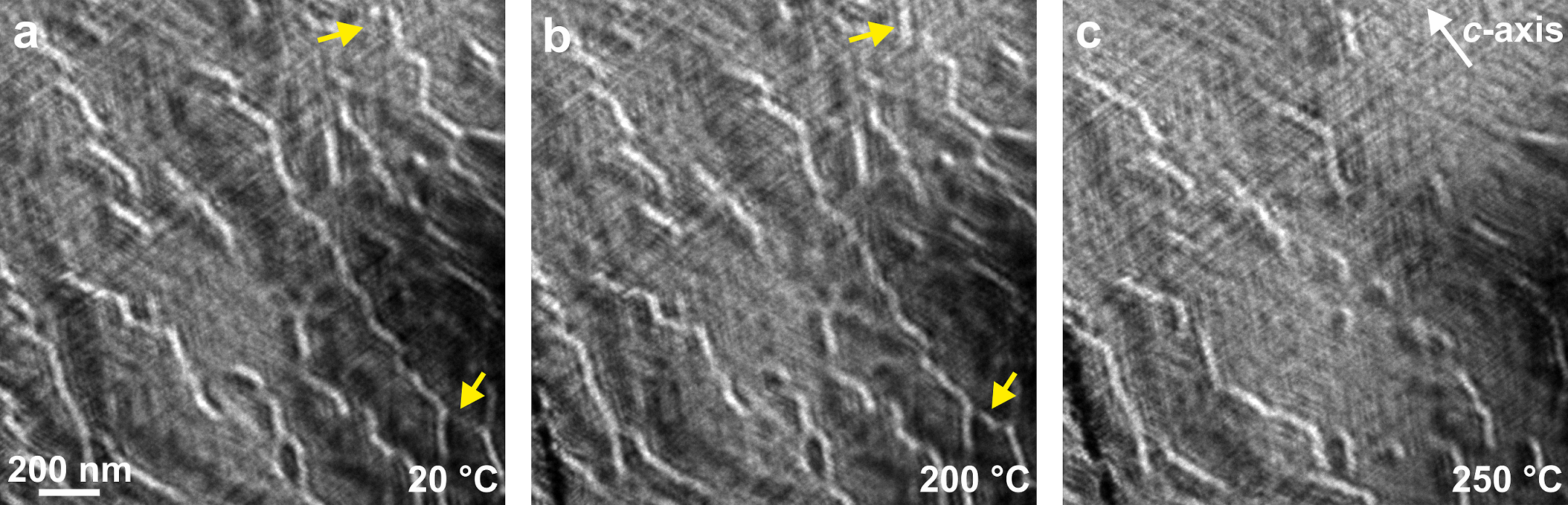}
\caption{\textbf{DW nucleation and annihilation.}  In-situ heating Fresnel images at 80 $\mu$m underfocus in an external magnetic field of 1.5 T perpendicular to the lamella at (\textbf{a}) 20 $^{\circ}$C, (\textbf{b}) 200 $^{\circ}$C and (\textbf{c}) 250 $^{\circ}$C, demonstrating DW nucleation (yellow arrows in panels a,b) and annihilation (panel c).}
\end{figure}

The observed magnetic textures and microstructural features are all nanoscopic, and in order to estimate their effect at the macroscopic scale the bulk magnetic properties of the samples were measured and are presented in Figure 6. All hysteresis loops have a shape typical for Sm(Co,Fe,Cu,Zr)$_z$ magnets, which does not change with temperature, as demonstrated for RT in Figure 6a and for 400 $^{\circ}$C in Figure 6b. When the external magnetic field is applied in the opposite direction to demagnetize the remanent state, there is an initial dip in the magnetization (between 0 and 0.5 T), hypothesized to be due to the Z phase reversing its magnetization before the rest of the magnet \cite{pierobon2020}. Figure 6c shows the temperature dependence of the saturation magnetization (solid symbols) and remanence (hollow symbols) for all three samples. Sm\textsuperscript{L}Cu\textsuperscript{H} has the highest saturation magnetization below 100 $^{\circ}$C, followed by Sm\textsuperscript{L}Cu\textsuperscript{L} and then Sm\textsuperscript{H}Cu\textsuperscript{H}. Above 100 $^{\circ}$C, Sm\textsuperscript{L}Cu\textsuperscript{L} surpasses Sm\textsuperscript{L}Cu\textsuperscript{H}. Quantitatively, the reduction in the saturation magnetization from RT to 400 $^{\circ}$C is 10.3\% for Sm\textsuperscript{L}Cu\textsuperscript{L}, 15.4\% for Sm\textsuperscript{L}Cu\textsuperscript{H} and 14.8\% for Sm\textsuperscript{H}Cu\textsuperscript{H}. Similarly, Sm\textsuperscript{L}Cu\textsuperscript{L} has the highest remanence, followed by Sm\textsuperscript{L}Cu\textsuperscript{H} and Sm\textsuperscript{H}Cu\textsuperscript{H} at all temperatures. Quantitatively, the reduction in the remanence from RT to 400 $^{\circ}$C is 16.4\% for Sm\textsuperscript{L}Cu\textsuperscript{L}, 15.9\% and Sm\textsuperscript{L}Cu\textsuperscript{H}, and 21.3\% for Sm\textsuperscript{H}Cu\textsuperscript{H}. Figures 6d-f show the dependence of the coercivity on temperature for each sample individually, revealing that at RT Sm\textsuperscript{L}Cu\textsuperscript{H} has the highest coercivity, followed by Sm\textsuperscript{L}Cu\textsuperscript{L} and Sm\textsuperscript{H}Cu\textsuperscript{H}. This changes at 400 $^{\circ}$C, where all magnets have approximately the same coercivity. Quantitatively, the coercivity reduction from RT to 400 $^{\circ}$C is 74.6\% for Sm\textsuperscript{L}Cu\textsuperscript{L}, 76.7\% for Sm\textsuperscript{L}Cu\textsuperscript{H} and 71.2\% for Sm\textsuperscript{H}Cu\textsuperscript{H}. The highest saturation magnetization and remanence and the lowest associated reduction with increasing temperature are expected for Sm\textsuperscript{L}Cu\textsuperscript{L}, because it has the highest amount of Co, an element with a high magnetic moment and Curie temperature. The opposite is valid for Sm\textsuperscript{H}Cu\textsuperscript{H}.

\begin{figure}[!htbp]
\centering
\includegraphics[width=0.95\columnwidth]{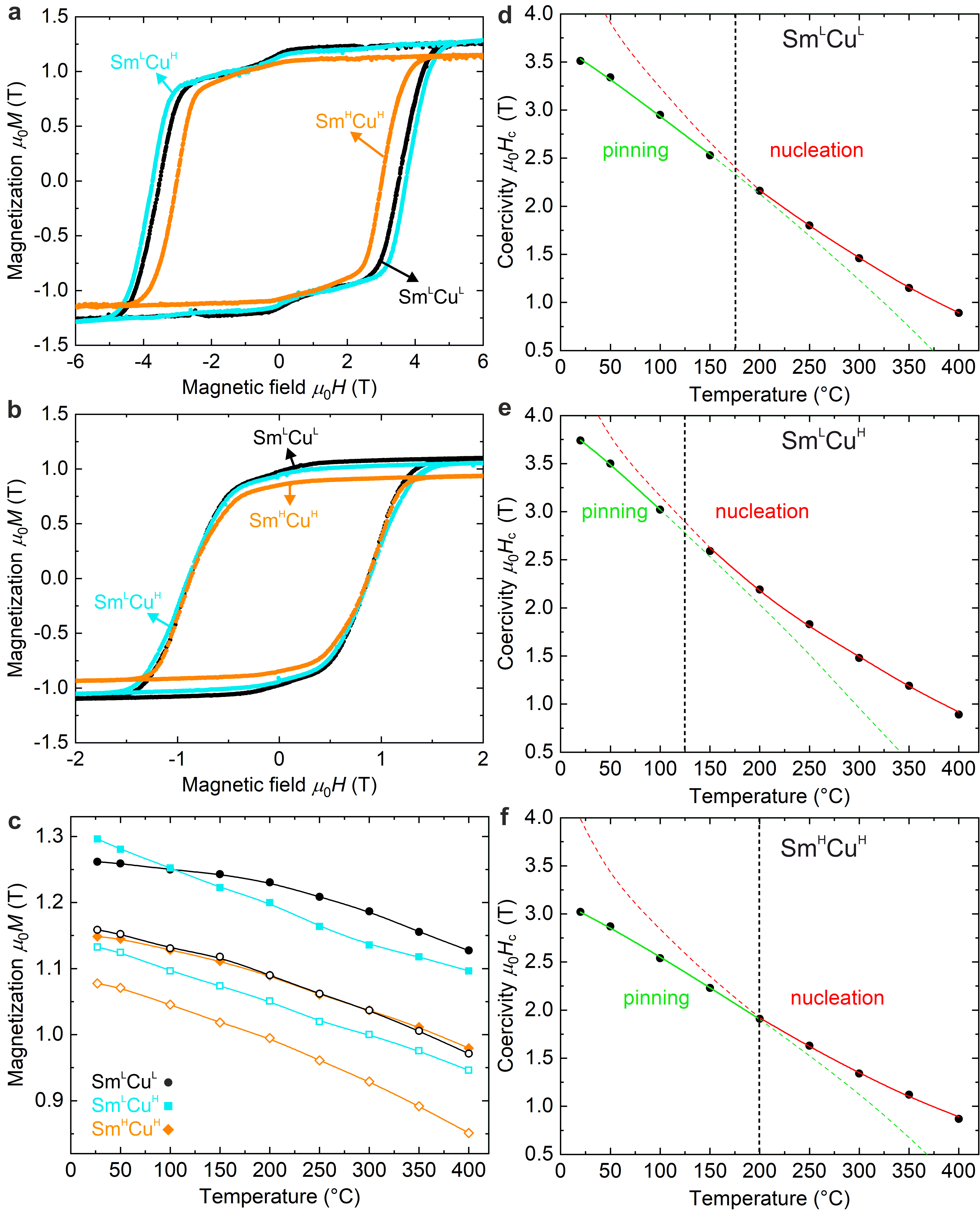} 
\caption{\textbf{Bulk magnetic properties.} Hysteresis loops of all samples at (\textbf{a}) RT and (\textbf{b}) 400 $^{\circ}$C having the same shape and revealing a magnetization dip before reversal. (\textbf{c}) Temperature dependence of the saturation magnetization (solid symbols) and remanence (hollow symbols) for all samples. Theoretical coercivity values for DW pinning (green) and nucleation (red) are fitted to the experimental values, showing that the transition happens around (\textbf{d}) 175 $^{\circ}$C for Sm\textsuperscript{L}Cu\textsuperscript{L}, (\textbf{e})  125 $^{\circ}$C for Sm\textsuperscript{L}Cu\textsuperscript{H} and (\textbf{f}) 200 $^{\circ}$C for Sm\textsuperscript{H}Cu\textsuperscript{H}.}
\end{figure}

The differences in the coercivity of the samples cannot be explained by considering only their bulk chemical compositions and microstructures. The magnetic structures and magnetization processes that arise from their microstructures also have to be considered. DW pinning at the cell walls occurs due to the difference in DW energy between the cells and cell walls. Our results show that Sm\textsuperscript{L}Cu\textsuperscript{H}, the sample with the largest Cu gradient, has the highest coercivity at RT, while Sm\textsuperscript{H}Cu\textsuperscript{H}, the sample with the smallest Cu gradient, has the lowest coercivity. Therefore, a Cu gradient across the cell walls enhances DW pinning by providing an additional DW energy difference, which increases the coercivity. Importantly, Sm\textsuperscript{H}Cu\textsuperscript{H} also has the thickest Z phase, which is another contributing factor to its relatively low coercivity. The influence of the Z phase is further described in the context of Supplementary Figure 1.

At high temperatures, thermal fluctuations overcome the energy barrier needed for magnetization reversal, leading to DW nucleation (as shown in Figure 5). In order to determine which coercivity mechanism is dominant, for each mechanism the theoretical values for the expected coercivity $H_{\rm c}$ are fitted to the experimental data in Figures 6d-f. For DW pinning, the following expression is used \cite{gaunt1972}:

\begin{align}
    H_{\rm c} = H_{0} - cT/M_{\rm s}
\end{align}

\noindent where $T$ is the temperature, $M_{\rm s}$ is the saturation magnetization, and the fitting parameters are $H_{0}$, representing the extrapolated absolute-zero coercivity, and $c$, describing the DW energy. Although $c$ is temperature dependent, for the purposes of differentiating between pinning and nucleation it is sufficient to consider it constant. For DW nucleation, the following expression is used \cite{givord1992}:

\begin{align}
    H_{\rm c} = aK_{1}/M_{\rm s} - nM_{\rm s}
\end{align}

\noindent where $K_{1}$ is the uniaxial magnetic anisotropy \cite{tellez1997}, and the fitting parameters are $a$, describing the defect type and dimensions, and $n$, representing the effect of dipolar interactions. Our fits demonstrate that the coercivity is dominated by DW pinning at low temperatures and DW nucleation at high temperatures. The transition between the two occurs between 150 and 200 $^{\circ}$C for Sm\textsuperscript{L}Cu\textsuperscript{L}, between 100 and 150 $^{\circ}$C for Sm\textsuperscript{L}Cu\textsuperscript{H}, and at 200 $^{\circ}$C for Sm\textsuperscript{H}Cu\textsuperscript{H}. 

It can be concluded that the sample with the highest Cu concentration has the lowest transition temperature, and vice versa. This can be explained by the fact that Cu segregates at the cell walls and reduces their Curie temperature, thus lowering the energy barrier for nucleation. Nucleation therefore happens at the cell walls and the transition temperature between pinning and nucleation decreases with increasing Cu concentration inside the cell walls, which in turn decreases the coercivity with increasing temperature. This invaluable data on the temperature behavior of Sm(Co,Fe,Cu,Zr)$_z$ magnets allows us to provide guidelines on how to directly improve their magnetic performance. Specifically, the Cu concentration should be carefully engineered so that the Cu gradient is as steep as possible to increase the coercivity while the total Cu concentration should be as low as possible to avoid the coercivity reduction with increasing temperature.

\section*{Conclusions}

By directly comparing atomic-resolution microstructural analysis and high-resolution magnetic imaging, we are able to identify key microstructural features that dominate the magnetic performance of composite Sm(Co,Fe,Cu,Zr)$_z$ magnets. The morphology of the SmCo$_5$ cell walls determines the magnetization texture. The denser the cell-wall network, the denser the DW network with more magnetization curling, which can lead to the formation of closed magnetic loops. Increasing the Cu gradient at the cell walls strengthens DW pinning at the cell walls and thus increases the coercivity. Contrarily, increasing the Z-phase thickness decreases the coercivity. The coercivity mechanism in Sm(Co,Fe,Cu,Zr)$_z$ is DW pinning at low temperatures and DW nucleation at high temperatures, and the transition between the two occurs between 100 $^{\circ}$C and 200 $^{\circ}$C. DW nucleation happens at the cell walls, and increasing the Cu concentration inside the cell walls decreases the transition temperature, which in turn increases the coercivity reduction with increasing temperature. From our results, we conclude that in order to improve the performance of Sm(Co,Fe,Cu,Zr)$_z$ magnets, the Z-phase thickness should be minimized and the Cu gradient at the cell-wall edges should be increased while keeping the total Cu concentration as low as possible. We provide a detailed understanding of the link between macroscopic magnetic properties and nanoscale microstructural and magnetic textures, and thus give precise directions on how the microstructure should be modified to further tailor the performance of Sm(Co,Fe,Cu,Zr)$_z$ magnets.

\section*{Methods}

\subsection*{Sample synthesis}

The alloying elements were molten in an induction furnace in an argon atmosphere of 99.999\% purity and cast in a metallic mold. The alloy was hammer-milled and then jet-milled to obtain a particle size of 4 -- 8 $\mu$m. The resulting powder was transferred to a rubber mold, where it was aligned with magnetic pulses of field strength 5 and then isostatically pressed at 300 MPa. This was sintered in vacuum at 1200--1220 $^\circ$C, solution-annealed at 1170--1200 $^\circ$C, and quenched to RT in an inert-gas atmosphere. The next steps were tempering at 850 $^\circ$C, slow-cooling to 400 $^\circ$C, and subsequent quenching to RT. All samples were produced by Arnold Magnetic Technologies. Table 1 shows the overall chemical composition of the materials used in this study.

\subsection*{Preparation of lamellae}

Electron-transparent lamellae for TEM measurements were made in a Helios 600i dual-beam focused ion beam (FIB) scanning electron microscope (SEM) workstation. Polished magnetic samples were Ga$^{+}$ sputtered and the lamellae were prepared using the conventional lift-out method. A Fischione Nanomill system was used for low-energy ($<$ 1 keV) Ar$^+$ milling, which significantly reduces ion-beam induced surface damage. 

\subsection*{Transmission electron microscopy}

The lamella thickness was measured using the EELS log-ratio technique on an FEI Tecnai F30 field-emission gun (FEG) transmission electron microscope.

HRSTEM EDX maps were obtained at an accelerating voltage of 200~kV, using an FEI Talos F200X microscope equipped with a Super-X EDS system.

The magnetic structure of the lamellae was investigated in their thermally demagnetized (magnetically pristine) state in zero magnetic field (Lorentz mode). The LTEM measurements were performed in the Fresnel mode, where electrons are deflected by the Lorentz force $\textbf{F}= -e\textbf{v}\times \textbf{B}$, with $e$ the electron charge, $\textbf{v}$ the velocity of electrons and $\textbf{B}$ the in-plane magnetic induction in the sample. EH measurements were performed using a magnetic biprism\cite{dunin2004} at a voltage between 90 and 100 V, which resulted in 3-nm fringe spacing and 75\% contrast. The room-temperature LTEM and EH images were obtained on a spherical aberration-corrected FEI Titan microscope at an accelerating voltage of 300~kV using a Gatan K2-IS direct-electron counting camera. The in-situ heating LTEM experiments were performed on the aforementioned FEI Tecnai F30 microscope. All images were recorded deploying the Gatan Microscopy Suite software.

Magnetization cycles and in-situ experiments were repeated several times for each sample to prove the reproducibility of the results.

\subsection*{Atom probe tomography}

The standard lift-out method on an FEI Helios Focused Ion Beam 600i workstation was used to prepare needle-shaped samples for APT analysis, which were mounted to a flat-top microtip coupon provided by Cameca. An apex of $<$70 nm diameter was achieved by applying sequential annular milling and low-kV cleaning, resulting in $<$0.01 at.\% Ga in the top 10 nm of the specimen. Data were collected between 5 kV and 9.5 kV on a LEAP4000X-HR atom probe microscope at 54 K by applying 100 pJ laser pulses at a frequency of 200 kHz, which give a Co charge-state ratio Co\textsuperscript{++}/Co\textsuperscript{+} between 5 and 10. These parameters and a chamber vacuum of 10\textsuperscript{-9} Pa resulted in a background level consistently below 20 ppm/ns. The atom-map reconstructions were validated by considering that the Z platelets are flat. Additionally, the lattice spacing was measured via spatial distribution maps along the easy axis, normal to the platelets, to validate the dimensional accuracy of the atom-map reconstruction.

\subsection*{Magnetometry}
A superconducting quantum interference device (SQUID), a part of Quantum Design's Magnetic Property Measurement System (MPMS3), was used to measure hysteresis loops along the easy axis of sample pieces with masses between 35 and 65 g at temperatures between RT and 400 $^{\circ}$C. 

\subsection*{Magnetic induction maps}

An electron wave passing through a magnetized sample gains an electromagnetic phase shift, as described by the Aharonov-Bohm effect \cite{ehrenberg1949,aharonov1959}:
\begin{align}
\varphi(x,y) & =\varphi_{\mathrm{el}}(x,y)+\varphi_{\mathrm{mag}}(x,y) = C_{\mathrm{el}}\int V(\boldsymbol{{\rm r}})dz-\frac{\pi}{\Phi_{0}}\int A_{z}(\boldsymbol{{\rm r}})dz,
\end{align}
with the electrostatic and magnetic contributions to the phase shift $\varphi_{\mathrm{el}}\left(x,y\right)$ in the electrostatic potential $V(\boldsymbol{{\rm r}})$ and $\varphi_{\mathrm{mag}}\left(x,y\right)$ in the vector potential $A_{z}(\boldsymbol{{\rm r}})$,
the magnetic flux quantum $\Phi_{0}=\pi\hbar/e$,
and the interaction constant $C_{\mathrm{el}}=\frac{\gamma m_{\mathrm{el}}e\lambda}{\hbar^{2}}$,
where $\gamma$ is the Lorentz factor and $\lambda$ is the electron wavelength. Here, the $z$-axis denotes the incident electron beam direction. The electrostatic contribution was assumed to be uniform throughout the sample, as compared to the magnetic contribution.

The magnetic induction maps shown in Figs. 3 and 4 were created in two steps. Firstly, the cosine of the magnetic phase $\varphi_{\mathrm{mag}}$ was calculated to create contours, so that the phase difference between two neighboring contours is 2$\pi$.  Secondly, the gradient of $\varphi_{\mathrm{mag}}$ was calculated to find the direction of the projected in-plane magnetic induction. This was visualized by superimposing a color scheme on the maps (as indicated by the color wheel in Figs. 5 and 6), where each color represents a specific direction. 

\bibliography{Paper2_bibliography}

\section*{Acknowledgements}

L.P., R.E.S., M.C., and J.F.L. gratefully acknowledge funding from the Swiss National Science Foundation (Grant No. 200021--172934). R.E.D.B. and A.K. gratefully acknowledge funding from the German Research Foundation (Project-ID 405553726 – TRR 270) and from the European Union’s Horizon 2020 Research and Innovation Programme (Grant No. 823717, project “ESTEEM3” and Grant No. 856538, project “3D MAGiC”). We also thank Joakim Reuteler for preparing the TEM lamellae, and ScopeM, ETH Zurich for the access to its facilities. 

\section*{Author contributions statement}

J.F.L., M.C. and R.E.D.B. initiated the study. J.F.L. supervised the work. U.V.W. provided the samples. L.P. and R.E.S. carried out the HRSTEM, in-situ LTEM, diffraction and E.E.L.S. experiments. S.S.A.G. performed the APT measurements. L.P. and A.K. carried out the room-temperature LTEM and EH experiments. A.F. performed the magnetic measurements. L.P. analyzed and correlated the data. All authors discussed the results and wrote the manuscript. 

\section*{Additional information}

The authors declare that they have no competing financial or non-financial interests.

\section*{Supplements}

The crystal orientation of all the lamellae in the study was the same in order to ensure a consistent comparison between the samples. Supplementary Figure 1a shows a high-resolution transmission electron microscopy (TEM) image of Sm\textsuperscript{L}Cu\textsuperscript{L}, illustrating Z-phase platelets (diagonal lines) in Sm$_2$Co$_{17}$ cells. The thickness of the platelets varies significantly. The cells below and above the bottom and the middle platelets are in a twinning relationship, as indicated by the yellow dashed lines following the direction of the atomic planes. This is further confirmed by the diffraction patterns recorded across, above and below the bottom platelet shown in Supplementary Figures 1b, c and d, respectively. The patterns above and below contain the reflections from [110] and [$\bar{1}\bar{1}$0] directions, respectively, while the pattern across contains the reflections from both. The twinning of the cells at the Z phase is an abrupt discontinuity in the crystal structure and the magnetocrystalline anisotropy. This may significantly reduce the coercivity of the magnet. 

\renewcommand{\figurename}{Supplementary Figure}

\begin{figure}[h]
  \centering
  \includegraphics[width=0.8\linewidth]{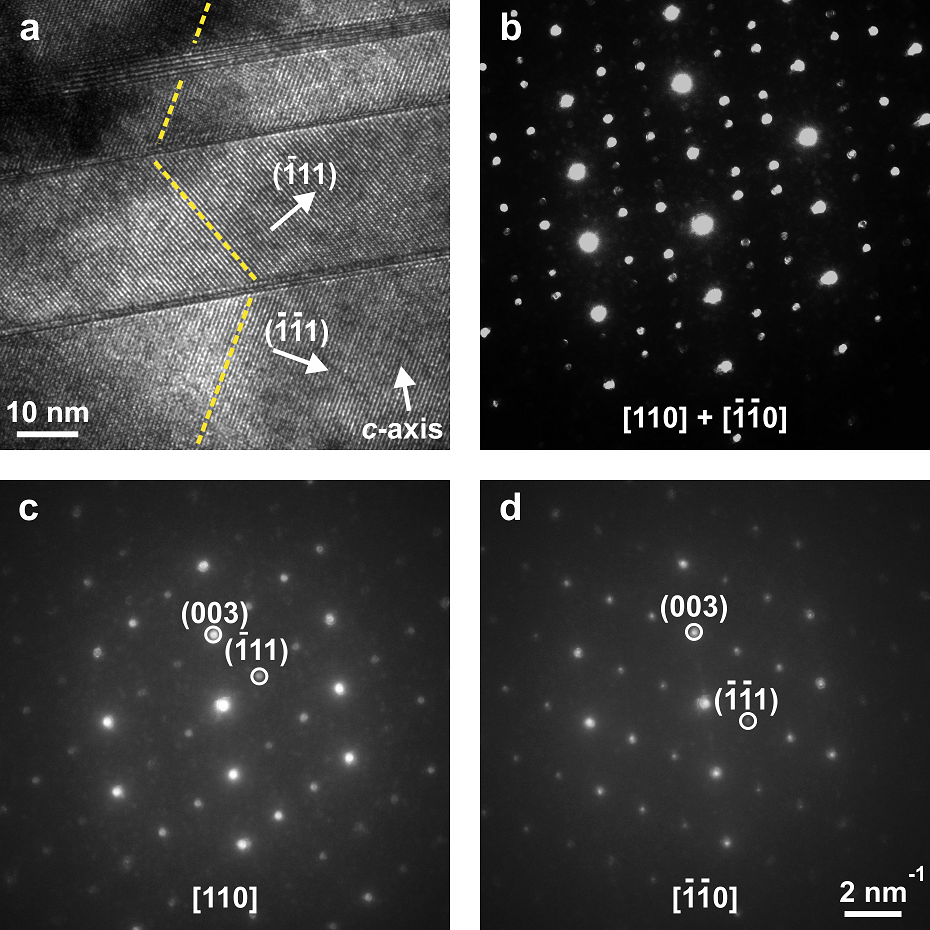}
  \caption{\textbf{Twinning in Sm(Co,Fe,Cu,Zr)$_z$.} (\textbf{a}) High-resolution TEM image of the Z-phase platelets in Sm$_2$Co$_{17}$ cells, with yellow dashed lines indicating twinning across the platelets. Diffraction patterns recorded (\textbf{b}) across, (\textbf{c}) above and (\textbf{d}) below the bottom platelet contain the reflections from [110] (panel c) and [$\bar{1}\bar{1}$0] (panel d) directions, confirming the twinning.}
\end{figure}

Atom probe tomography (APT) was used to obtain proxigrams (proximity histograms) of individual elements in the Z phase. Proxigrams show the concentration of elements as a function of distance from a predefined interface. In this case, the predefined interface corresponds to all surfaces with a Zr concentration of 5\%, which are the edges of the Z phase. As shown in Supplementary Figure 2, the Zr distribution in all samples is almost identical, with the Zr concentration peaking at 19.3 $\pm$ 0.2\% in Sm\textsuperscript{L}Cu\textsuperscript{L}, 18.7 $\pm$ 0.4\% in Sm\textsuperscript{L}Cu\textsuperscript{H} and 19.2 $\pm$ 0.2\% in Sm\textsuperscript{H}Cu\textsuperscript{H}.

\newpage

\begin{figure}[h]
  \centering
  \includegraphics[width=1\linewidth]{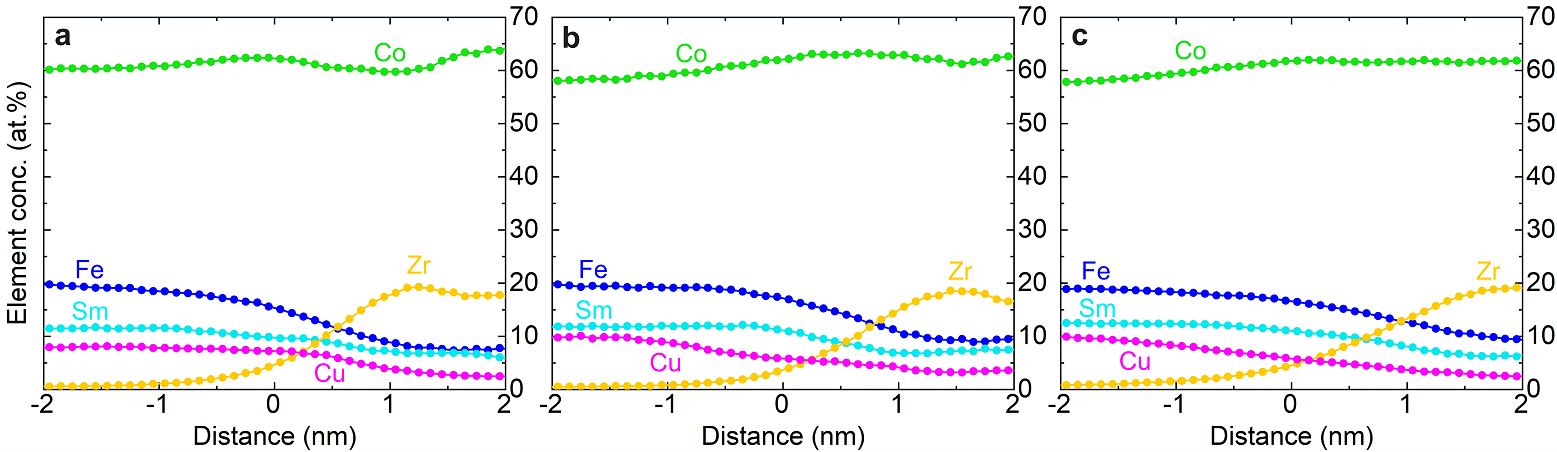}
  \caption{\textbf{APT measurements of the Z phase.} Proxigrams of individual elements with respect to the edges of the Z phase for (\textbf{a}) Sm\textsuperscript{L}Cu\textsuperscript{L}, (\textbf{b}) Sm\textsuperscript{L}Cu\textsuperscript{H} and (\textbf{c}) Sm\textsuperscript{H}Cu\textsuperscript{H}, revealing that the element distribution across the Z phase is almost identical in all samples.}
\end{figure}

The magnetic structure of the samples was imaged in the Fresnel mode at 80 $\mu$m underfocus in an external magnetic field of 1.5 T applied by the objective lens perpendicular to the lamella (only perpendicular fields could be applied by the objective lens). Supplementary Video 1 shows for Sm\textsuperscript{L}Cu\textsuperscript{H} the change in the magnetic texture from 20 to 400 $^{\circ}$C. Specifically, the domain walls (DWs) indicated in Figures 5a,b in the main text are nucleated at 150 and 200 $^{\circ}$C. The coercivity decreases with increasing the temperature, so the magnetization becomes saturated and DWs are annihilated by the external magnetic field above 200 $^{\circ}$C, although some DWs survive up to 300 $^{\circ}$C. When the sample is cooled back to 20 $^{\circ}$C, it stays saturated perpendicular to the lamella (parallel to the external magnetic field).

\renewcommand{\figurename}{Supplementary Video}
\setcounter{figure}{0}

\begin{figure}[h]
\centering
  \includegraphics[width=0.5\linewidth]{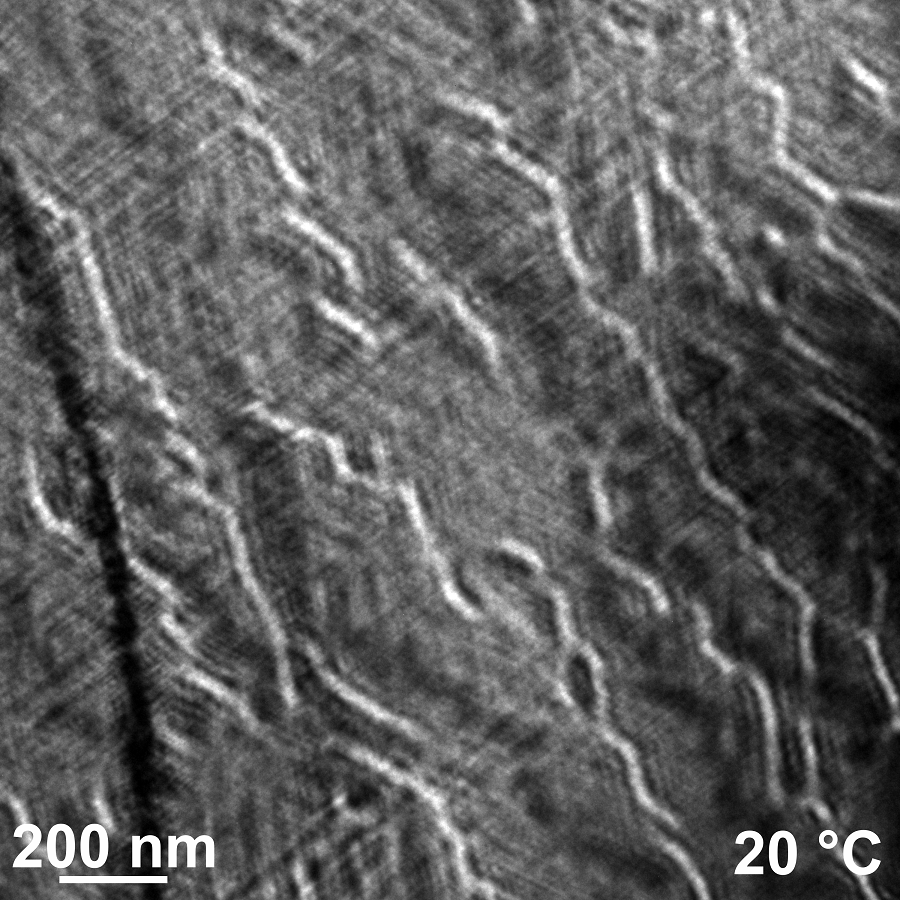}
  \caption{\textbf{In-situ Lorentz TEM.} Fresnel images of the magnetic structure of Sm\textsuperscript{L}Cu\textsuperscript{H} at 80 $\mu$m underfocus in an external magnetic field of 1.5 T perpendicular to the lamella, showing DW nucleation at 150 and 200 $^{\circ}$C and annihilation above 200 $^{\circ}$C. The video is accessible online.}
\end{figure}

\end{document}